# Analytical Classification of Multimedia Index Structures by Using a Partitioning Method-Based Framework

Mohammadreza keyvanpour [1] and Najva Izadpanah [2]

[1]Department of Computer Engineering, Alzahra University, Tehran, Iran
`keyvanpour@Alzahra.ac.ir`
[2]Department of Computer Engineering, Islamic Azad University, Qazvin Branch, Qazvin,Iran
`n.izadpanah@qiau.ac.ir`

## ABSTRACT

Due to the advances in hardware technology and increase in production of multimedia data in many applications, during the last decades, multimedia databases have become increasingly important. Content-based multimedia retrieval is one of an important research area in the field of multimedia databases. Lots of research on this field has led to proposition of different kinds of index structures to support fast and efficient similarity search to retrieve multimedia data from these databases. Due to variety and plenty of proposed index structures, we suggest a systematic framework based on partitioning method used in these structures to classify multimedia index structures, and then we evaluated these structures based on important functional measures. We hope this proposed framework will lead to empirical and technical comparison of multimedia index structures and development of more efficient structures at future.

## 1. INTRODUCTION

In the recent years, due to Popularity of multimedia data including text, image, audio, video and graphic object, and also increase in production of such multimedia data in many applications, it is necessary to develop efficient methods for storing and retrieval of multimedia data.

In past years researchers have tried extending retrieval techniques used in text retrieval to the area of multimedia retrieval. In text-based multimedia retrieval approach, a set of keywords are assigned to each multimedia object. However, there are some important limitations to text-based multimedia retrieval approach. Since in text-based approach each multimedia object needs to be manually annotated with keywords, it is impractical for large data sets. And, due to the subjectivity of the human annotator, the annotations may not be complete. Also, it may be infeasible to describe visual content of multimedia objects using keywords [1, 2].

Motivated by the lack of an efficient multimedia retrieval technique, content-based multimedia retrieval was introduced. Content-based multimedia retrieval is one of the most interesting research areas in multimedia database field. This approach is used to retrieve desired multimedia object from a large multimedia database. In content-based multimedia retrieval approach, low-level features like colour, object motion and shape automatically extracted from the multimedia objects and mapped to feature vectors. Therefore, multimedia objects are represented as a vector of extracted visual features [2].

A content-based image retrieval system performs multimedia retrieval by computing similarities between multimedia objects and the multimedia query, and the results are ranked based on the computed similarity values. The similarity of two features is a function of their distance in the multidimensional space.

In fact, in content-based multimedia retrieval, at search phase, retrieval system retrieves the most similar objects to query object by performing similarity comparison between query vector and all the object vectors in database [2, 3].

For having a fast and efficient similarity search, it is necessary to store the multimedia feature vectors in a multidimensional index structure to support fast and efficient search [4, 5]. Multidimensional index structures are data structures specifically designed for the storage and management of multidimensional data. In the literature, multidimensional index structures are also referred to as multidimensional access methods [6]. In this study by multimedia index structures we refer to multidimensional index structures for support content-based multimedia retrieval systems.

Lots of research on content-based multimedia retrieval systems has led to many kinds of multimedia index structures. The main goal of these structures is to improve the performance in the retrieval of multimedia data that satisfy a search query. Two most common query types that must be solved by multimedia index structures are point query and range query [4, 6].

Most of multimedia index structures have a tree-like structure. In these structures, data points are stored in data nodes (leaves) and each directory node points to a set of sub trees. There is a single directory node, which is called the root [4].

Due to variety and plenty of proposed multimedia index structures in past decades, we suggest a systematic framework based on partitioning method used in these structures to classify multimedia index structures, and then we evaluated these structures based on important functional measures.

The rest of the paper is organized as follows: In section 2, we review related work on classification of multimedia Index structures. We describe multimedia content-based retrieval in section 3. In section 4, the proposed partitioning method-based framework for classify multimedia index structures is presented. Then in section 5, we evaluate multimedia index structures. And, section 6 includes the conclusions.

## 2. RELATED WORK

During past decades, different kinds of index structures for supporting efficient similarity search in multidimensional databases including multimedia databases have been proposed.

In the literature multidimensional index structures (multidimensional access methods) totally have been classified in to two classes: Point access methods (PAMs) and Spatial access methods (SAMs) [6].

In other hand, some researches classified multidimensional index structures based on either they construct in the feature space (such as k-d-b-tree [7], R-tree [8]) or construct in metric space (such as M-tree [9], VP-tree [10]) [4].

PAMs have been designed to perform searches on point databases that store only points that do not have a spatial extension [6]. Points are organized in a number of buckets which in PAMs corresponds to some subspace of the data space. As in [11] PAMs has classified by the properties of the bucket regions: structures with disjoint regions( such as grid file [12], quad-tree [13],k-d-b-tree, and buddy tree [11])or structures with mutual overlapping regions (such as twin grid file [14]), structures that may have the region shape of box ( such as grid file, quad-tree, k-d-b-tree, twin grid file , multilevel grid file [15], and buddy tree)or structures that have some arbitrary polyhedral shape regions( such as BSP-tree[16], BANG file [17]),structures that cover the complete data space (such as grid file, quad-tree, k-d-b-tree, twin grid file, BSP-tree, and BANG file) or structures that just cover those parts that contain data points(such as multilevel grid file and buddy tree).

SAMs have been developed lately to manage extended objects, such as lines, polygons. As in [6, 18] SAMs has classified in to five categories: methods that transforms objects either into a higher dimensional space and then support them by PAMS (like what is done in LSD-tree [19]) or in to one dimensional space by Space-Filling Curves and then support by one dimensional structures such as B-tree (like what is done in ZKDB+-tree [20]),structures with overlapping regions(R-tree, R*-tree [21], SR-tree [22],).Clipping-based methods that do not allow any overlaps between bucket regions (R+-tree [23]), and Multiple Layers that allow data regions of different layers overlap (multi-layer grid file [24], R-file [25]).

In this study, in order to present a comprehensive classification and evaluation of multimedia index structures, we extend classification that proposed in [26], and classified multimedia index structures based on the partitioning method used in these structures.

## 3. Multimedia content-based retrieval

Multimedia data refers to several kinds of data types. The basic types of data that characterize multimedia data are including text, images, audio, video and graphic objects. They have been used in a wide variety of applications such as: computer-aided design, geographical information systems, medical image management systems, and etc [2, 5].

Due to large amount of multimedia data that produce in many applications, it is necessary to develop efficient multimedia data management system for storing and retrieval of multimedia data. Motivated by this need, content-based retrieval systems was introduced. Content- based means that the technology makes direct use of content of the multimedia rather than relying on human annotation of data with keywords [1, 2].

The Basic architecture of Content-based multimedia retrieval systems is shown in Figure 1. It consists of three principal components: Feature Extraction, Index Structure and Retrieval Engine.

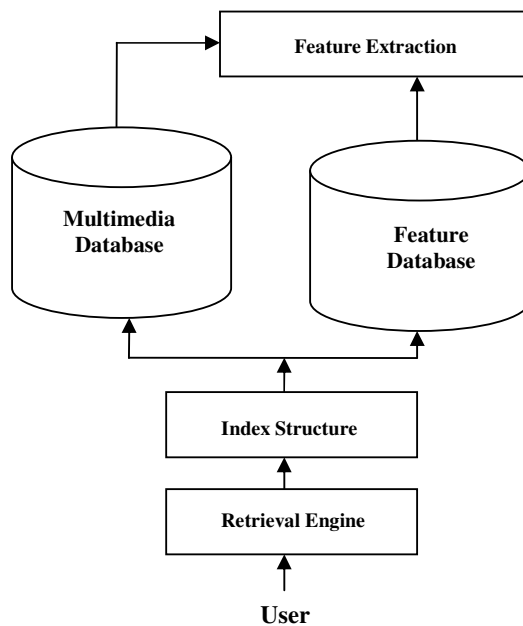

Figure 1. Basic architecture of Content-based Multimedia Retrieval systems

Feature Extraction: in content-based retrieval approach, low-level features from each multimedia object and mapped to multidimensional feature vectors. Therefore, multimedia object is represented as a vector of extracted low-level features [2].

Index Structure: As size of data is very large, index structures used to store data points to accelerate search operation [4].

Retrieval Engine: at search phase, retrieval engine retrieves the most similar objects to query object by performing similarity comparison between query vector and all the object vectors in database [3, 4].

## 4. PARTITIONING METHOD-BASED FRAMEWORK FOR CLASSIFY MULTIMEDIA INDEX STRUCTURES

This section includes the proposed framework for classify multimedia index structures. According to our study on multimedia index structures, here, we have classified them based on partitioning method used in these structures. Our classification of multimedia index structures is shown in Figure 2.

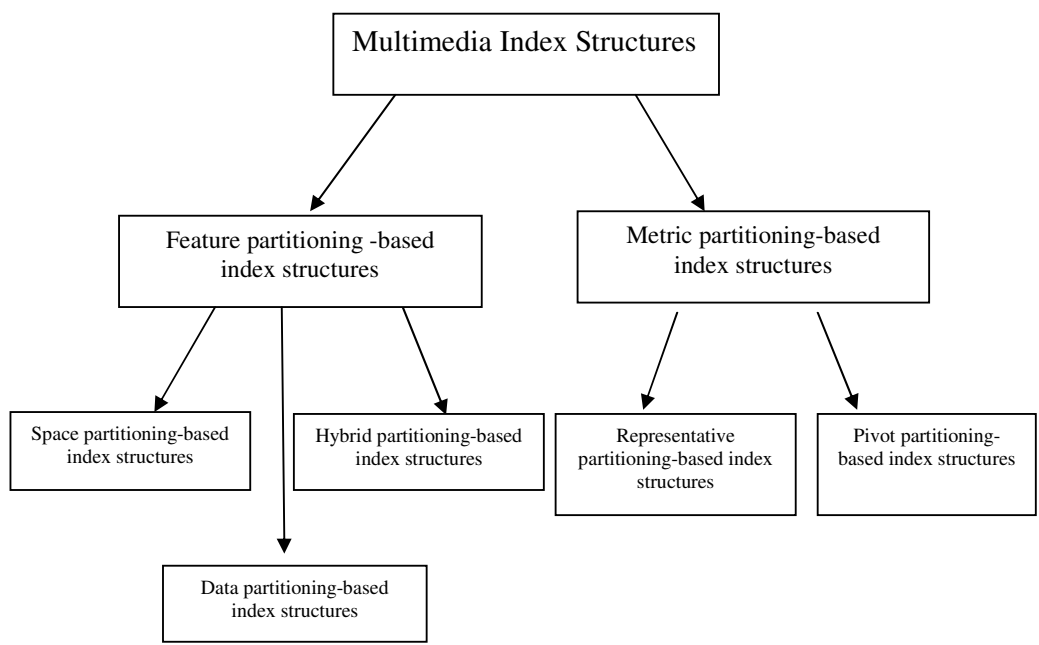

Figure 2. Classification of multimedia index structures

Many multimedia index structures have been proposed to improve the multimedia image databases performance. An index structure is first characterized by its page capacity that characterized in (1) [27].

$$P_{capacity}=|D|/|P| \qquad (1)$$

Where, $P_{capacity}$ is page capacity, $|D|$ is data size which is proportional to the dimension of the data space, and $|P|$ is page size which is fixed by the file system.

An Index structure is also characterized by its selectivity, the selectivity of an indexing structure characterizes its efficiency in terms of number of secondary memory accesses during search.

More specifically, it defines the average proportion of file pages that are actually pruned during the search. The selectivity characterized in (2) [27].

$$S = \left(1 - \frac{FC_{eff}}{|DB|}\right) \qquad (2)$$

Where, $S$ is selectivity, $C_{eff}$ is efficient capacity that it characterize by the average number of data points per page, $F$ is the number of page read, and |DB| is database size.

### 4.1. Feature Partitioning-Based Index Structures

Feature partitioning-based index structures include structures that partition feature vector space to build an index structure in order to retrieve similar data points to the query object.

As in [26] Feature partitioning-based index structures has classified into two categories: space partitioning-based indexing structures, and data partitioning-based indexing structures. Lately, researchers combined positive aspects of space partitioning-based structures and data partitioning-based structures to build more efficient index structures [28]. Therefore as it is shown in Figure 2, we classified feature partitioning-based index structures in to three categories: Space partitioning-based index structures, data partitioning-based index structures, and hybrid partitioning-based index structures.

#### 4.1.1. Space Partitioning-Based Index Structures

Space partitioning-based index structures partition the data space into sub regions by an iso-oriented hyper plane, passing through a data vector. Each sub region is according to a disk page, and as soon as each sub region becomes full, that region is splitting in to sub regions. Some examples of these structures are Grid file, K-D-B-tree, Quad-tree.

An example of space partitioning-based index structures is shown in Figure 3.

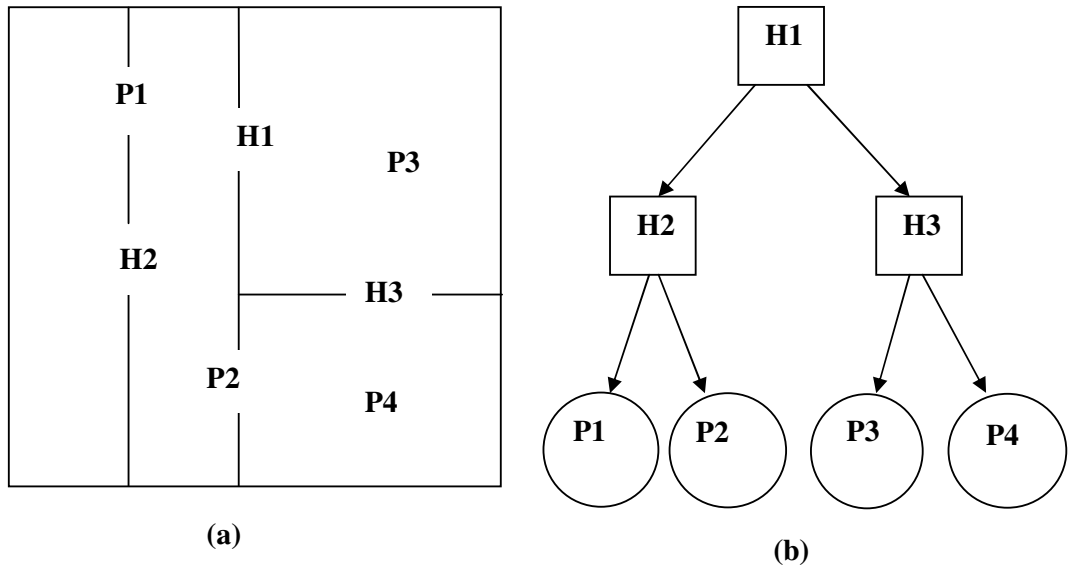

Figure 3.  An example of space partitioning-based index structures. (a):  partitioning of space by iso-oriented hyper planes (H1, H2, and H3) in to partitions p1,p2, p3, and p4. (b):  Index structure according to (a).

Since these structures partition whole of space into hyper rectangles, depends on the structure of data distribution, there may be many empty spaces with out any data points in sub regions that lead to poor storage utilization of these structures [4, 28].

### 4.1.2. Data Partitioning-Based Index Structures

Data partitioning-based index structures partition the database into overlapping regions that enclosed either by a MBR (Minimum Bounding Rectangle), or by a hyper-sphere. Some researches uses intersection of MBRs and hyper-spheres as enclosing regions [4, 6].

The idea of data partitioning-based index structures is to use overlapping regions to improve storage utilization, but overlapping between directory regions slows down the search. Due to growth of overlaps between directory regions in high-dimensional spaces, selectivity of these methods in high-dimensional spaces degrades [6, 27].

Some main examples of data partitioning-based index structures are R-tree, SS-tree, SR-tree , and X-tree [29].

An example of partitioning of data space by minimum bounding rectangles in R-tree and the corresponding R-tree structure is shown in Figure 4.

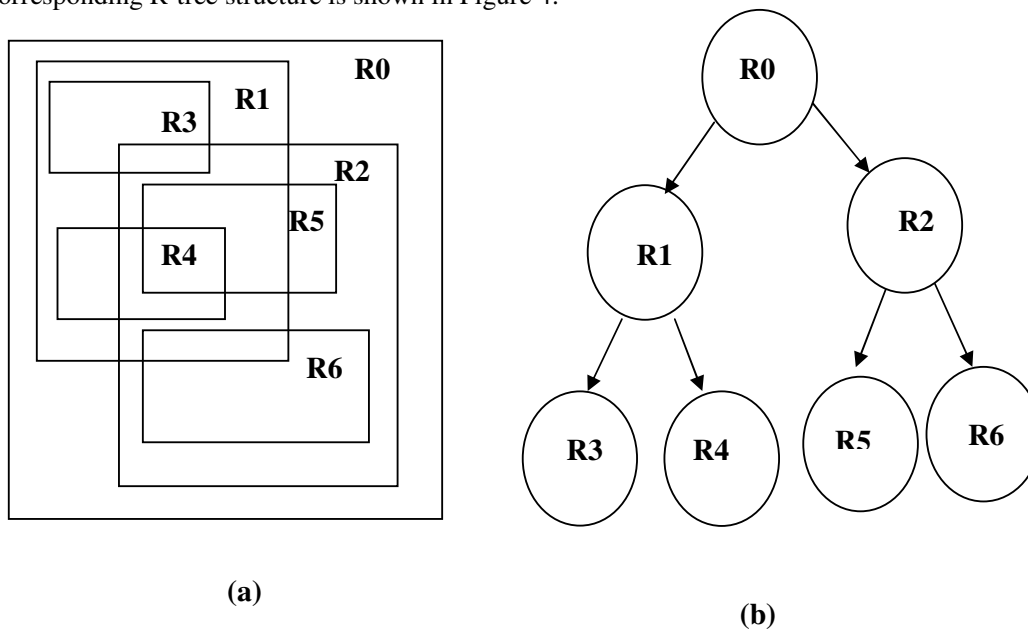

(a)

(b)

Figure 4. An example of R-tree structure. (a): partitioning of data space R0 in to overlapping regions R1, R2, R3, R4, R5, and R6. (b): R-tree structure according to (a).

### 4.1.3. Hybrid Partitioning-Based Index Structures

In this study, Hybrid partitioning-based index structures refers to those structures that take advantages of both space partitioning-based and data partitioning-based index structures and combines positive aspects of them in a single structure to improve selectivity.

Hybrid partitioning-based index structures recursively partitions data space into overlapping sub regions, but to guarantee storage utilization and reduce side effect of overlapping between sub regions, in these structures, at each sub region uses bounding forms such as MBRs to enclosed data points and sub regions are allowed overlapping if splitting in to sub regions can guarantee storage utilization. Some examples of these structures are Hybrid-tree [28], and SH-tree [30].

An example of partitioning of data space by hybrid partitioning-based index structures that uses MBR as bounding forms is shown in Figure 5.

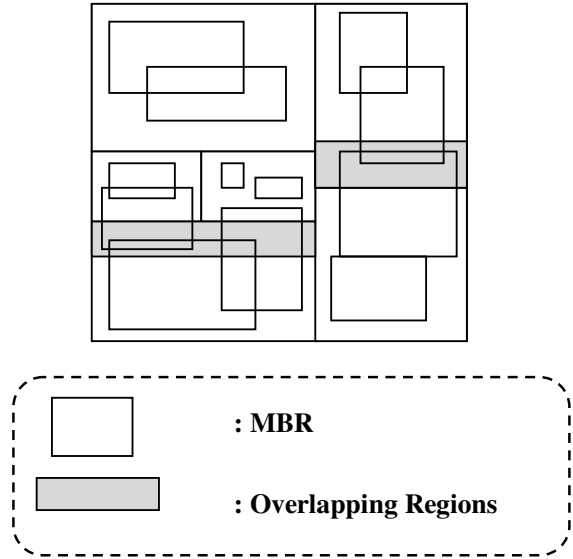

Figure 5. An example of partitioning of data space by a hybrid partitioning-based index structure that uses MBR as bounding forms

### 4.2. Metric Partitioning-Based Index Structures

Metric partitioning-based index structures unlike feature partitioning-based index structures are built only based on the relative distances between the data points instead of their absolute positions in a feature space.

The main aim of metric partitioning-based index structures is indexing in metric spaces to retrieve data objects which are similar to a query object.

A metric space is defined in (3) [9].

$$M = (D, d) \qquad (3)$$

Where, D is a domain of feature values, and d is a distance function with three properties includes symmetry, non negativity, and triangle inequality.

Symmetry, non negativity and triangle inequality properties are defined in (4), (5), and (6) respectively.

$$d(O_x, O_y) = d(O_y, O_x) \qquad (4)$$

$$d(O_x, O_y) > 0, (O_x \neq O_y), \text{ and } d(O_x, O_x) \qquad (5)$$

$$d(O_x, O_y) \leq d(O_x, O_z) + d(O_z, O_y) \qquad (6)$$

Where, Ox, Oy are two data objects, d is distance function.

Researches on metric partitioning-based index structures have been shown that these structures totally have good performance in case of selectivity and storage utilization. And the applicability of these structures is not limited to feature vector spaces [9, 27].

In this study, as it is shown in Figure 2, we classified metric partitioning-based index structures in to two categories: representative partitioning-based index structures, and pivot partitioning-based index structures.

### 4.2.1. Representative Partitioning-Based Index Structures

Representative partitioning-based index structures partition the dataset by selecting the representatives of data points to build an index structure.

In each level, a set of m representative is selected and other data points assign to the nearest representative. The distance from the representative to the farthest associated data point is said covering radius. A directory node saves representatives at each level. Partitioning is continuing until a maximum number of points per leaf are obtained [4, 27].

Examples of representative partitioning-based index structures are M-tree [9], and Slim-tree [31].

An example of metric space partitioning by representatives is shown in Figure 6.

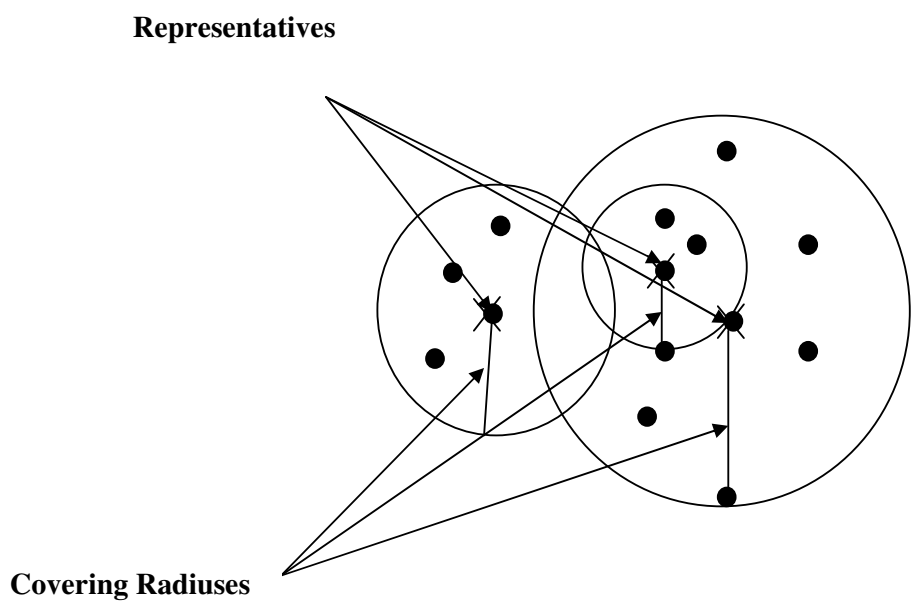

Figure 6. An example of metric space partitioning by representatives

### 4.2.2. Pivot Partitioning-Based Index Structures

Pivot partitioning-based index structures use pivot (vantage) points to partition the dataset in order to build an index structure.

At each level one or more Pivot points are selected and the data set is partitioned in two regions base on median distance to the vantage points.

62

Since pivot partitioning-based index structures need all data points in the database, they can't be updated dynamically [27].

Examples of representative partitioning-based index structures are ‹VP-tree [10], Multiple-VP-tree [32], and VP Forest [33].

An example of metric space partitioning by pivot point is shown in Figure 7.

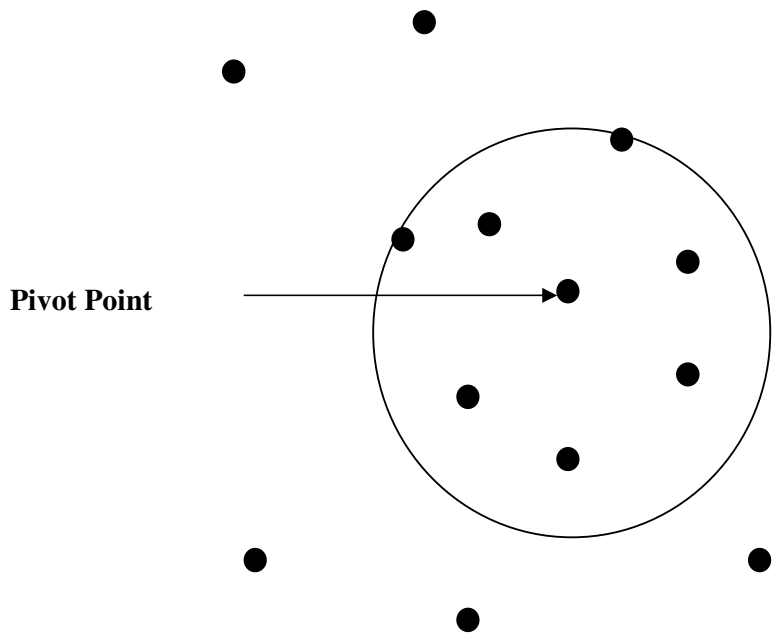

Figure 7. An example of metric space partitioning by pivot point

## 5. Evaluation of Multimedia Index Structures

In this section, we evaluate multimedia index structures according to main functional measures. Our evaluation is summarized in Table 1.

The functional measures that considered in our evaluation of multimedia index structures are as follows:

- Selectivity: a multimedia index structures must provide low number of file access during the search and then high selectivity is guaranteed.
- Scalability: a multimedia index structures must provide high scalability in dealing with high number of data and high number of dimensionality.
- Dynamicity: a multimedia index structures must have a dynamic structure. It means that structure incrementally constructed as the data points are inserted in the database.
- Storage utilization: a multimedia index structures must guarantee storage utilization.
- Minimum overlapping between nodes: a multimedia index structures must guarantee minimum overlapping between nodes.

## 6. CONCLUSIONS

Due to impressive increase in importance of multimedia data in many applications, and importance of designing optimum index structures in fast and efficient content-based retrieval of multimedia data from multimedia databases, it is necessary to classify and evaluate multimedia index structures.

Our study on multimedia index structures shows that multimedia index structures can classified based on the partitioning method used in these strutures in to five main classes: space partitioning-based index structures, data partitioning-based index structures, hybrid partitioning-based index structures, representative partitioning-based index structures, pivot partitioning-based index structures. Three firt of these classes including structures that construct in feature spaces, and two last classes including structures that construct in metric spaces.

In this study we proposed some functional measures according to this classification to evaluate multimedia index structures. And then we evaluated multimedia index structures according to that proposed functional measures.

The proposed classification framework is comprehensive and could guide researchers to develop more efficient methods in this field.

For future work, we plan to extend our classification framework for each five classes of multimedia index structures.

Table 1. Evaluation of multimedia index structures

| Multimedia index structures | | Functional measures | | | | |
|---|---|---|---|---|---|---|
| | | *Selectivity* | *Scalability* | *Dynamicity* | *Storage utilization* | *Minimum overlapping between nodes* |
| Feature partitioning - based index structures | Space partitioning-based index structures | Medium | Poor | Good | Poor | Good |
| | Data partitioning-based index structures | Medium | Poor | Good | Good | Poor |
| | Hybrid partitioning-based index structures | Good | Good | Good | Good | Medium |
| Metric partitioning-based index structures | Representative partitioning-based index structures | Good | Good | Good | Good | Poor |
| | Pivot partitioning-based index structures | Good | Medium | Poor | Good | Good |

## REFERENCES

[1]     K. Chakrabarti , (2001) *Managing Large Multidimensional Datasets Inside a Database System*, PhD   Thesis, University of Illinois at Urbana-Champaign. Urbana, Illinois.

[2]      D. Feng, W.C. Siu, H.J. Zhang, (2003)" Multimedia information retrieval and    management: Technological fundamentals and applications", Berlin: Springer.

[3]      Ch. Srinivasa Rao, S. Srinivas Kumar, and B. Chandra Mohan, (2010)" Content based Image Retrieval Using Exact Legendre Moments And Support Vector Machine" ,The international journal of Multimedia & Its applications(IJMA),VOL.2,NO.2.

[4]     C. BÖHM, S. BERCHTOLD,and D. A. KEIM, (2001) "searching in high-dimensional spaces-Index structure for Improving the performance of Multimedia Databases", ACM Computing surveys, vol.33, No.3, pp.322-373.

[5]     M.S.LEW, N.SEBE, C.DJERABA, and R. JAIN, (2006)"Content-based Multimedia Information Retrieval: State of the Art and Challenges", ACM Transactions on Multimedia Computing, Communications, and Applications.

[6]     V. Gaede, O. Günther, (1998)" Multidimensional Access Methods", ACM Computing Surveys, Vol. 30, No. 2.

[7]     J. T. Robinson, (1981)"The K-D-B-tree: A search structure for large multidimensional dynamic indexes", In Proc. ACM SIGMOD Int. Conf. on Management of Data, pp. 10-18.

[8]     A. Guttman,(1984) "R-trees: A Dynamic Index Structure for Spatial Searching", In Proc. ACM SIGMOD Int. Conf. on Management of Data, Boston, MA, pp. 47-57.

[9]     P. Ciaccia, M. Patella, and P. Zezula,(1997)" M-tree: An efficient access method for similarity search in metric spaces", In VLDB '97: Proc.23rd International Conference onVery Large Data Bases, pages 426–435, San Francisco, CA, USA.

[10]    P. N. Yianilos, (1993)" Data structures and algorithms for nearest neighbor search in general metric spaces", In SODA '93: Proc. fourth annual ACM-SIAM Symposium on Discretealgorithms, pages 311–321, Philadelphia, PA, USA.

[11]    B.Seeger, B. and H.-P. Kriegel, (1990)"The buddy-tree: An efficient and robust access method for spatial data base systems", In Proc 16th Int. Conf. on Very Large Data Bases, pp. 590-601.

[12]    J.Nievergelt, H. Hinterberger, and K. C. Sevcik, (1984)"The grid file: An adaptable, symmetric multi key file structure", ACM Trans. Database Systems 9 (1), 38-71.

[13]    H. Samet, (1984) "The quad tree and related hierarchical data structure", ACM Computing Surveys 16 (2), 187-260.

[14]    A.Hutesz, H.-W. Six and P. Widmayer, (1988)"Twin grid files: Space optimizing access schemes", In Proc. ACM SIGMOD Int. Conf. on Management of Data, pp.183-190.

[15]    K.-Y .Whang, and R. Krishnamurthy (1985), "Multilevel grid files", Yorktown Heights, NY: IBM Research Laboratory.

[16]    H.Fuchs, Z.Kedem, and B.Naylor(1980)," On visible surface generation by a priori tree structures", Computer Graphics 14 (3).

[17]    M. Freeston, (1987)" The BANG file: A new kind of grid file", In Proc. ACM SIGMOD Int. Conf. on Management of Data, pp. 260-269.

[18]    H.-P. Kriegel, and B. Seeger, (1988)" PLOP-hashing: A grid file without directory", In Proc. 4th IEEE Int. Conf. on Data Eng., pp. 369-376.

[19]    A.Henrich, H.-W. Six, and P. Widmayer, (1989) " The LSD tree: Spatial access to multidimensional point and non-point objects", In Proc.15th Int. Conf. on Very Large Data Bases, pp. 45-53.

[20]    J.A. Orenstein,(1986) "Spatial query processing in an object-oriented database system", In Proc. ACM SIGMOD Int. Conf. on Management of Data, pp. 326-333.

[21]    N.Beckmann, H.-P. Kriegel,R. Schneider, and B. Seeger, (1990)."The R*-tree: An efficient and robust access method for points and rectangles", In Proc. ACM SIG-MOD Int. Conf. on Management of Data, pp. 322-331.

[22]    N. Katayama N., S. Satoh, (1997) "The SR-tree: An Index Structure for High-Dimensional Nearest Neighbor Queries", Proc. ACM SIGMOD Int. Conf. on Management of Data, pp.369-380.

[23]    T.Sellis, N. Roussopoulos, and C. Faloutsos, (1987)" The R+-tree: A dynamic index for multi-dimensional objects", In Proc. 13th Int. Conf. on Very Large Data Bases, pp. 507-518.

[24]     H.Six, and P. Widmayer,(1988)" Spatial searching in geometric databases", In Proc.4th IEEE Int. Conf. on Data Eng., pp. 496-503.

[25]     A.Hutesz, H.-W. Six, and P. Widmayer ,(1990)"The R-file: An efficient access structure for proximity queries", In Proc. 6th IEEE Int. Conf. on Data Eng., pp.372-379.

[26]     C. Li, E. Chang, H. Garcia-Molina, and G. Wiederhold, (2002) " Clustering for Approximate Similarity Search in High-dimensional Spaces", IEEE Transactions on Knowledge and Data Engineering,pp. 792 – 808.

[27]     N. Mo¨enne-Loccoz. Computer Vision Group, Computing Science Center ,(2005)*High-Dimensional Access Methods for Efficient Similarity Queries*, University of Geneva24 rue du G´en´eral Dufour, CH - 1211 Geneva 4, Switzerland.

[28]     K. Chakrabarti, S.Mehrotra, (1999)"The hybrid tree: an index structure for high dimensional feature spaces", Proc.IEEE international confererance on data engineering.

[29]     S. Berchtold., D. Keim, H.-P Kriegel,(1996)"The X-Tree: An Index Structure for High-Dimensional Data", 22nd Conf. on Very Large Databases, Bombay, India, pp. 28-39.

[30]     D. T. Khanh,(2006)" The SH-Tree: A Novel and Flexible SuperHybrid Index Structure for Similarity Search on Multidimensional Data", International Journal of Computer Science & Applications Vol. III, No. I1 – 25.

[31]     J. Caetano Traina, A. J. M. Traina, B. Seeger, and C. Faloutsos,(2000)" Slim-trees: High performance metric trees minimizing overlap between nodes", In Proc.7thInternational Conference on Extending Database Technology, pages 51–65, London, UK, .Springer-Verlag.

[32]     T.Bozkaya, Ozsoyoglu M,(1997)" Distance-based indexing for high-dimensional metric spaces*",* Proc. ACM SIGMOD International Conference on Management of Data, SigmodRecord Vol. 26, No. 2, pp. 357-368.

[33]     P. Yianilos , (1999)"Excluded middle vantage point forests for nearest neighbor search*",* In Proc.DIMACS Implementation Challenge, ALENEX`99.

**Authors**

Mohammadreza Keyvanpour is an Associate Professor at Alzahra University, Tehran, Iran. He received his B.s in software engineering from Iran University of Science &Technology, Tehran, Iran. He received his M.s and PhD in software engineering from Tarbiat Modares University, Tehran, Iran. His research interests include image retrieval and data mining.

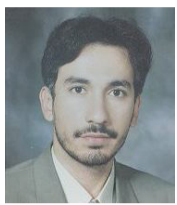

Najva Izadpanah received her B.s in software engineering from University of Science & Culture,Tehran, Iran. Currently, she is pursuing M.s in software engineering at Islamic Azad University, Qazvin Branch, Qazvin, Iran. Her research interests include image indexing and data mining.

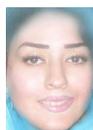